\documentstyle[times,epsf,subeqnarray,jprart,journals]{aipproc}

\newcommand{\Bc}{B_{\rm c}}

\begin{document}

\title{Hadronic blazar models and correlated X-ray/TeV flares}

\author{J\"org P. Rachen}

\address{Sterrenkundig Instituut, Universiteit Utrecht, 3508 TA Utrecht, The
Netherlands}

\maketitle

\vspace*{-20pt}

\begin{abstract}
The hypothesis that AGN jets might be the sources of the ultra-high energy
cosmic rays has originally motivated the venture of TeV gamma ray
astronomy. Surprisingly, after the discovery of TeV emission from blazars the
attention has shifted to more traditional explanations which do not involve
energetic hadrons, and there is even common believe that a hadronic
interpretation is disfavored by observations. It is shown here that this is
not the case, and that the currently observed spectra and variability
features of blazars can be perfectly understood within hadronic blazar
models. I also discuss how hadronic models might be observationally
distinguished from common leptonic models, and point out some interesting
aspects which could be relevant for the understanding of the differences
between blazar classes.
\end{abstract}

\section*{Why hadronic models?}
\vspace{-5pt}
\subsubsection*{AGN jets and the origin of cosmic rays}
\vspace{-5pt}

Cosmic rays are observed up to the enormous energy of $3\mal10^{20}\eV$, and
both theoretical and observational arguments suggest an extragalactic origin
of these most energetic particles \cite{Bie97jphG}. Since the cosmic ray
arrival directions are largely randomized by Galactic and extragalactic
magnetic fields, their sources cannot be easily identified from direct
observations. However, if we assume that they gain their energy by
acceleration (rather than by quantum processes, i.e., the decay of superheavy
particles) their extreme maximum energy allows to derive quite restrictive
selection criteria for possible acceleration sites. Most acceleration
scenarios in astrophysics, for example Fermi acceleration
\cite{Dru83rpp}, assume that particles are magnetically confined for
some time $t_{\rm acc}$ in the accelerating region. 
This implies three fundamental constraints on the maximum
proton energy,
\eqns{Emax}
\subeq{conf} E_{\rm cr} &<& e B R \Gamma\;, \\
\subeq{sync} E_{\rm cr} &<& \TS \frac32 m_p c^2 \Gamma 
	(m_p/m_e) \sqrt{\eta\Bc/\alpha_{\rm f} B}\\
\subeq{GZK} E_{\rm cr} &<& \eta e B \lambda_{\rm GZK} \Gamma,
\text
which we call (a) the {\em confinement limit}, (b) the {\em synchrotron
limit}, and (c) the {\em GZK limit}. The confinement limit simply expresses
that the particle gyro-radius $r_{\rm L}$ in the magnetic field $B$ is
smaller than the system size $R$. The synchrotron limit expresses that
$t_{\rm acc}$ is smaller than the synchrotron loss time, where $B_c \approx
4 \mal 10^{13}\G$ is the critical magnetic field, and $\alpha_{\rm
f}\approx 1/137$ the fine structure constant. The parameter $\eta = \Delta
E/E$ is the fractional energy gain in a Larmor time, $t_{\rm L} = E/e B c$,
and one can show that for Fermi acceleration generally $\eta<1$ (see
\cite[App.\,D]{RM98prd} for some discussion and references). The GZK limit
expresses that proton acceleration must be faster than photohadronic losses
at the microwave background, called the Greisen-Zatsepin-Kuzmin or GZK effect
\cite{GZK66}, and $\lambda_{\rm GZK} \sim 10\Mpc$ is the appropriate
attenuation length of protons with $E>10^{20}\eV$ \cite{Ste68prl}. The GZK
limit dominates over the confinement limit for sources on supercluster scales
(${\gsim}10\Mpc$), but it applies also to somewhat smaller scales
(e.g. clusters of galaxies) since in such objects usually $\eta\ll 1$
\cite{LSSacc}.

Only very few cosmic sources have been found which may satisfy all three
conditions for $E_{\rm cr}\gsim 10^{20}\eV$, and all of them are connected to
strong shocks in relativistic flows ($\Gamma\sim 10{-}1000$): (a) the
termination shocks of extended jets in radio galaxies \cite{RB93aa}, (b)
compact jets in blazars \cite{Man93aa,MPR99prd}, and (c) internal or external
shocks in fireballs proposed to induce Gamma-Ray Bursts \cite{GRBCR95}. All
of them are known as powerful emitters of non-thermal photons, in particular
gamma rays. If we adopt the assumption that there is a universal ratio
between cosmic ray and non-thermal photon emission (which is rather
simplistic, but not unreasonable as an estimate), we could use their total
contribution to the observed extragalactic gamma rays as a scale of their
total power as cosmic ray sources. This clearly favors AGN over GRB, because
AGN are known to produce $10{-}30\%$ of all extragalactic gamma-rays
(counting both resolved sources and diffuse background), while resolved GRB
contribute less than $1\%$. Due to the GZK effect, another important
selection criterion for UHECR sources is proximity.  This disfavors powerful
quasars, and also GRB if their cosmological distribution follows the star
forming activity, since in both cases most power is emitted at large
redshifts. In contrast, BL Lac objects and their proposed unbeamed
counterpart, FR-I radio galaxies \cite{UPad95pasp}, seem to be more frequent
in the current epoch than at large redshifts \cite{BBD+98aa}. Since their
jets may also satisfy the condition for the acceleration of UHECR, and they
contribute significantly to the non-thermal radiation in the universe,
blazars and FR-I radio galaxies may be regarded the primary candidates for
the origin of the highest energy cosmic rays.

\subsubsection*{Hadronic vs. leptonic gamma-ray emission in blazars}
\vspace*{-5pt}

Energetic hadrons can lead to the emission of gamma rays via $pp$
interactions with surrounding gas, or $p\gamma$ interactions with ambient
photons. This leads to the production of secondary $e^\pm$ pairs, or mesons
like $\pi^\pm$ and $\pi^0$, which eventually decay into $e^\pm$ pairs,
photons and neutrinos. Electrons or pairs can produce high energy photons by
synchrotron (or Compton) processes. The photons can either escape from the
jet or produce new pairs in $\gamma\gamma\to e^+e^-$ processes, which
subsequently radiate a new generation of photons. In particular {\em
synchrotron-pair cascades} of this kind are important if interactions of
UHECR protons in AGN jets are considered, where they shift the energy in
secondary radiation down from the extreme proton energies to the observable
gamma-ray regime. This mechanism has been coined PIC for ``proton induced
cascades'' \cite{MKB91aa}.

The term ``hadronic blazar models'' is used for a large variety of models for
the gamma-ray production in blazars, not all of which involve UHE cosmic
rays. All in common is just that they propose energetic protons as the main
carrier of dissipated energy in the jet, rather than so-called ``leptonic
models'', which assume the bulk of the energy available for radiation in
electrons or $e^\pm$ pairs. Some hadronic models assume gamma-ray production
in $pp$ interactions invoked in the collision of jets with surrounding gas
clouds, or in a very massive jet itself \cite{pp-junk}. In a different class
of models, protons have been suggested as being responsible just for the
injection of energetic electrons, which then produce the observed photon
emission by a synchrotron-self Compton (SSC) mechanism
\cite{Apostolos}. These models offer an interesting explanation for the
observed strong variability in gamma-ray blazars due to intrinsic
instabilities. They usually require only moderate proton energies, but very
large densities of relativistic protons in the jet.
 
Following the motivation given above, I want to focus here on a different
kind of hadronic models in which UHE cosmic rays in the jet interact with low
energy target photons. They are split into two classes: (a) target photons
are produced by synchrotron-radiating electrons co-accelerated with the
protons \cite{Man93aa}, and (b) external target photons are present in the
vicinity of the jet, as for example thermal photons emitted from an accretion
disk or a warm dust torus \cite{Pro97iau}. In a realistic scenario, photons
from both sources would be present. The local ratio of the external and
internal photon energy density in a reference frame comoving with the jet
differs hereby from the the ratio of the observed external and internal
luminosities by a factor $\Gamma_{\rm jet}^6$ (the corresponding number ratio
of external to internal photons scales with $\Gamma_{\rm jet}^4$).  It is
therefore hard to constrain the local comoving density of external photons in
the jet by observations. However, an indirect constraint on the external
photon density has been pointed out by Protheroe and Biermann \cite{PB97app}:
using a common accretion disk/torus model to estimate the radiation fields in
AGN, they show that the emission of $\TeV$ photons would be strongly
suppressed by $\gamma\gamma$ absorption, if the emission region is close
enough to the AGN that external photons could be relevant for $p\gamma$
interactions. This is inconsistent with the observed spectra in TeV blazars
like Mrk\,501 or Mrk\,421 (see below). Nevertheless, external photons can
still be important in high luminosity EGRET blazars (like 3C\,279)
\cite{MPR99prd}.

For the discussion of TeV blazars, we can therefore concentrate on so-called
{\em synchrotron-self proton induced cascade} (SS-PIC) models, in which UHE
protons interact with synchrotron photons emitted by electrons accelerated
{\em in the same process} as the protons. One clear consequence of such
models is that the gamma-ray emission due to the PIC process is in
competition with SSC emission of the electrons. The relation of the
luminosities of both processes is expressed by
\eqn{PIC:SSC}
\frac{L_{\rm PIC}}{L_{\rm SSC}} \lsim 0.1 \frac{u_p(\hat E_p)}{u_{\rm ph}}
	\left[\frac{B}{1\G}\right]^3 \left[\frac{R}{10^{16}\cm}\right]^2\;,   
\text 
where $u_p = N_p^2 dN_p/dE_p$ is the energy density of protons at the maximum
energy, $\hat E_p$, and $u_{\rm ph}$ is the bolometric energy density of the
target photons (i.e., the {\em synchrotron} photons emitted by primary
electrons), and $B$ and $R$ are magnetic field and size of the emission
region, respectively. The relation gives in fact only an upper limit on the
PIC contribution, since approximate equality assumes $\hat E_p = e B R$. We
see that, if the UHE proton content of the jet is significant ($u_p(\hat E_p)
\ge u_{\rm ph}$), and the maximum energy sufficient to explain the observed
UHECR ($e B R \Gamma_{\rm jet} \gsim 10^{20}\eV$, assuming a jet Lorentz
factor $\Gamma_{\rm jet}\sim 10$), PIC emission will dominate over SSC at
compact jet scales ($R\lsim 10^{16}\cm$) --- and vice versa. For the likely
case of adiabatic scaling of a dominantly transversal magnetic field with the
jet radius, $B\propto R^{-1}$, the relative contribution of SSC emission
would increase on larger scales. Therefore both processes may contribute to
the observed radiation, although emission which is variable on short time
scales (requiring small $R$) would tend to be dominated by PIC. Defining
clear signatures to distinguish PIC from SSC emission can therefore make the
hypothesis that blazars and FR-I radio galaxies are the sources of UHE cosmic
rays testable by GeV-TeV gamma-ray observations.
 
\section*{Spectral properties of SS-PIC models}
\vspace{-5pt}
\subsubsection*{Synchrotron-pair cascades in power law target spectra}
\vspace{-5pt}

At frequencies below the X-ray band, blazar spectra are well represented by
broken power laws, in some cases even by single power laws down to the sub-mm
regime \cite{Sambruna}. These photons play a triple role in SS-PIC models:
(i) they are targets for the initial $p\gamma$ interactions, (ii) they are
targets for the propagation of synchrotron pair cascades, and (iii) they
allow clues on the particle spectrum of the primary accelerator. Let us, for
simplicity, assume that the soft target photon number spectrum is described
by a single power law, $dN_{\rm ph}/dE \propto E^{-\alpha_t-1}$,
$\alpha_t\sim 1$ is called the energy index of the target spectrum. The
stationary electron spectrum producing these photons by synchrotron radiation
must then have the form $dN_e/dE\propto Q_{e,\rm acc} t_{\rm syn} \propto
E^{-\alpha_e-1}$ with an energy index $\alpha_e = 2\alpha_t$, where $Q_{\rm
acc}\propto E^{-\alpha_{\rm acc}-1}$ is the number of electrons injected by
the accelerator per unit time, and $t_{\rm syn}\propto E^{-1}$ is the
synchrotron cooling time of the electrons. This means that the spectrum
injected by the accelerator is a power law with energy index $\alpha_{\rm
acc} = \alpha_e-1$. In contrast to electrons, cooling of protons is usually
dominated by adiabatic \cite{RM98prd} or advection losses
\cite{Man93aa}. Hence, the proton cooling time $\bar t_p$ can be considered
as energy independent, and the stationary proton spectrum is $dN_p/dE\propto
Q_{p,\rm acc}(E)$. If the accelerated protons have the same spectrum as the
electrons, the proton power law energy index is $\alpha_p \propto \alpha_{\rm
acc} = 2\alpha_t-1$, but we discuss also different choices of $\alpha_p$.

The opacity for the photohadronic production is $\tau_{p\gamma} = \bar
t_p/t_{p\gamma}$, where $t_{p\gamma} \propto E^{\alpha_t}$ is the according
loss time scale. The major channel for hadronic gamma production is $p\gamma
\to \pi^0 + \ldots$ with the subsequent decay $\pi^0\to\gamma\gamma$. The
resulting photons are dominantly absorbed by the soft target photons through
$\gamma\gamma$ pair production. The $\gamma\gamma$ opacity has the same
energy dependence as $\tau_{p\gamma}$ for constant $\bar t_p$, that is
$\tau_{\gamma\gamma} \propto E^{-\alpha_t}$, hence the stationary energy
distribution of the primary photons from pion decay is $dN_\gamma^{[0]}/dE =
dN_p/dE (\tau_{\gamma\gamma}/\tau_{p\gamma}) \propto dN_p/dE$. Saturated
production of pairs in $\gamma\gamma$ collisions
(i.e. $\tau_{\gamma\gamma}\gg 1$) then leads to \cite{Sve87mnras}
\eqn{pairspec}
\frac{dN_\pm^{[1]}}{dx} \propto \frac1{x^2}\int_{2x}^{\hat x_0} 
dx'\tau_{\gamma\gamma}(x') \frac{dN_\gamma^{[0]}}{dx'} 
\propto x^{-\alpha_\pm-1} 
= \left\{\begin{array}{l@{\for}l} 
	x^{\alpha_t-\alpha_p-2} & \alpha_p > \alpha_t \\  
	x^{-2}			& \alpha_p \le \alpha_t 
	\end{array}\right.  
\text
where $x$ is the energy of photons or pairs in units of $m_e c^2$, and $\hat
x_0 = \hat \gamma_p m_\pi/2m_e \sim 100\hat \gamma_p$ is the maximum $x$ of
the primary injected photons. The pairs produce a new generation of photons
by synchrotron radiation, which are distributed in the stationary, saturated
case as 
\eqn{photspec}
\frac{dN_\gamma^{[1]}}{dx_1} 
\propto \frac{Q_{\rm syn}^{[1]}(x_1)}{\tau_{\gamma\gamma}(x_1)} 
\propto x_1^{-\alpha_t-3/2}
	\left[\frac{x^2 dN_\pm(x)}{dx}\right]_{x=x_1^{1/2}} \;,
\text
i.e., as a power law with energy index $\alpha_1 = \frac12\alpha_\pm +
\alpha_t$. $Q_{\rm syn}^{[1]}(x)\propto x^{-\alpha_\pm/2}$ is the injected
number of synchrotron photons per unit time at energy $m_e c^2 x$ by the
first generation of pairs. The stationary photons can inject a second
generation of pairs, and so on. This {\em synchrotron-pair cascade}
continues, and with each step the characteristic photon energy is reduced by
\eqn{cascade:x}
x_{n} \approx \frac{B}{4 B_c} x_{n-1}^2 \for B x_{n-1} < 4 B_c\;.
\text
The latter condition expresses the classical limit of synchrotron radiation
and is mostly fulfilled in hadronic blazar models; in the non-classical case
the cascade propagates approximately as $x_n = \frac12 x_{n-1}$. Thus, the
energy is rapidly reduced in each step once $x\ll B_c/B\sim 10^{12}$, and
since $\tau_{\gamma\gamma}\propto x^{\alpha_t}$ the
opacity for the production of subsequent cascade generations quickly
decreases. The cascade becomes unsaturated at a photon energy ${\sim}m_e c^2
x_{\gamma\gamma}$, defined by $\tau_{\gamma\gamma}(x_{\gamma\gamma}) = 1$,
and dies out rapidly for $x<x_{\gamma\gamma}$. The photon spectrum emerging
from the emitter is \cite{Sve87mnras,MKB91aa}
\eqn{spectrum}
\frac{dN_{\gamma,\rm em}}{dx} 
\propto \sum_{n=1}^{n^* +1} Q_{\rm syn}^{[n]}(x) 
	\left[\frac{1-e^{-\tau_{\gamma\gamma}(x)}}
		{\tau_{\gamma\gamma}(x)}\right]
\propto \left\{\begin{array}{l@{\for}l} 
	x^{\alpha_t-\alpha_1-1} & x \ll x_{\gamma\gamma}\\
	x^{-\alpha_1-1}         & x \gg x_{\gamma\gamma}
	\end{array}\right.\quad,  
\text
where the sum extends over all cascade generations and $n^*$ is the last
generation with a highest photon energy $\hat x_{n^*}> x_{\gamma\gamma}$
($\hat x_n$ is determined by repeated application of \eq{cascade:x} on $\hat
x_0$). It can be shown that for typical blazar conditions $n^* = 3{-}4$, and
that $Q_{\rm syn}^{[n^*]}$ dominates the emission around $x_{\gamma\gamma}$
\cite{Man93prd}. 

The term in brackets in \eq{spectrum} has the meaning of a mean escape
probability of the photons from the emission region, and is strictly correct
only for a plane-parallel geometry. However, the asymptotic behavior of the
spectrum would be the same in any geometry, that is, $x_{\gamma\gamma}$ marks
a break in the spectrum by $\Delta\alpha = \alpha_t$, which we call the {\em
opacity break}. This broken power law shape is typical for photospheric
emission, which is in the nature of the SS-PIC models where the soft photons
are responsible both for the production and absorption of gamma rays. It is
directly observable unless there is significant absorption by external
photons surrounding the jet, which would lead to an exponential cutoff
$\propto \exp(-\tau_{\gamma\gamma,\rm ext})$. The observed $\GeV{-}\TeV$
spectra of TeV-blazars support in most cases a broken power law shape, as
expected from the SS-PIC model for $m_e c^2 x_{\gamma\gamma} \Gamma_{\rm
jet}\sim 1\TeV$, but are inconsistent with a rapid cutoff in this regime
\cite{Sambruna}.  This has been used as an argument against external PIC
models for TeV blazars \cite{PB97app}. An observed opacity break in the TeV
regime is also expected from a direct determination of $\tau_{\gamma\gamma}$
for typical blazar parameters \cite{Man93aa}, and from general considerations
concerning the efficiency of hadronic blazar models \cite{MPR99prd}.

\subsubsection*{Robust features of hadronic blazar spectra}
\vspace{-5pt}

If we ignore the details of the cascade spectra in the vicinity of cutoffs or
breaks, the spectral propagation can be described in terms of a simple
algebra. If we call $\alpha_n(x_n)$ the local spectral index in the vicinity
$x\sim x_n$, we can introduce the relations
\eqns{operators}
\subeq{+} \alpha_n(x_n) =& f_+[\alpha_{n-1}(x_{n-1})] 
		&\for x_n > x_{\gamma\gamma} \\   
\subeq{-} \alpha_n(x_n) =& f_-[\alpha_{n-1}(x_{n-1})] 
		&\for x_n < x_{\gamma\gamma}   
\text
with
\conteqno\eqns{operators}
\subeq{def} f_\pm[\alpha_n] &=& 
	\max\left\{\TS\frac12(\alpha_n\pm\alpha_t+1),\frac12\right\}\quad,
\text
where the dimensionless photon energy $x_n$ propagates following
\eq{cascade:x}. The spectrum of the $k$-th cascade is then given by
$f_+^k[\alpha_p]$ above the opacity break at $x_{\gamma\gamma}$, and by
$f_-[f_+^{k-1}[\alpha_p]]$ below (but above $x'_{\gamma\gamma} = B
x_{\gamma\gamma}^2/4 B_c \ll x_{\gamma\gamma}$), where $f_+^k$ denotes the
$k$-fold iterative application of $f_+$. 

If we ask which cascade generation is dominant at a given energy we encounter
a problem related to the energy propagation equation \refeq{cascade:x}: for
any given primary spectral feature at $x_0$, the position of its ``image'' in
the $n$-th cascade is dependent on the magnetic field in the emitter by
$x_n\propto B^{2^n-1}$. This means, for example, that the cutoff of the
\mbox{4-th} cascade generation is $\hat x_4 \propto B^{15}$, so that a
variation of $B$ by a factor of $2$ over the emitting region would ``smear
out'' the value of $\hat x_4$ by more than 4 orders of
magnitude. Fortunately, it is still possible to make some quite robust
predictions on hadronic blazar spectra, since the cascade generation spectra
converge quickly with $k$ to $f_+^\infty[\alpha_p] = 1+\alpha_t$, and
$f_-[f_+^\infty[\alpha_p]]= 1$, independent of $\alpha_p$. Moreover, the
power contained in each cascade generation is approximately equal for $n<
n^*+1$ and rapidly decreases for larger $n$, so that the spectrum below
$x_{\gamma\gamma}$ is not strongly changed by adding generations with
$n>n^*$. This allows the statement that {\em hadronic gamma-ray blazar
spectra are described by a broken power law with energy indices
$\alpha\approx 1$ below, and $\alpha\approx 1 +\alpha_t$ above a break
observed at ${\sim}1\TeV$, where $\alpha_t$ is the IR-X energy index of the
source}.

So far we did not consider the cutoff in the target photon spectrum, $\hat
x_t$, which is generally observed at an energy $m_e c^2 \hat x_t \Gamma_{\rm
jet}$ between in the optical and X-ray regimes. We consider the target photon
spectrum for $x_t>\hat x_t$ as a power law with a steep energy index,
$\alpha_t' > \alpha_t + 1$. Gamma rays with $x_n < \hat x_t^{-1}$ then induce
cascade photons below $\tilde x = B/(4 B_c \hat x_t^2)$ with an index
$\alpha'_{n+1} \approx f_-'[f_+^\infty[\alpha_p]] = \frac12$, where $f_-'$ is
the cascade operator defined in \eq{operators:def} replacing $\alpha_t$ with
$\alpha_t'$ (for $\tilde x < x_{\gamma\gamma}$). Since this result in
independent of the detailed value of $\alpha_t'$ as long as
$\alpha_t'>\alpha_t+1$, we can consider it as valid for any steep cutoff,
e.g., an exponential one. The energy $m_e c^2 \tilde x \Gamma_{\rm jet}$
marks a break in the spectrum, which is observable above the cutoff of the
primary electron emission ($\tilde x > \hat x_t$) if
\eqn{tcut-break:cond}
\hat \eps_t = m_e c^2 \Gamma_{\rm jet} \hat x_t 
	\lsim 100\eV \left[\frac{\Gamma_{\rm jet}}{10}\right]
		\left[\frac{B}{\G}\right]^{1/3}\;.
\text
For $B<100\G$ this transition corresponds to the observationally motivated
distinction between so-called low-energy-cutoff blazars (LBLs) and
high-energy-cutoff blazars (HBLs) \cite{UPad95pasp}. In the SS-PIC models we
therefore expect fundamentally different shapes for the high energy spectra
of these classes, as illustrated in Figure \reffig{spectra}. Obviously, the
spectrum for LBLs can easily be confused with an SSC spectrum, while a clear
difference is predicted for HBLs in the MeV gamma-ray regime.
\setfig{t}{spectra}
\epsfysize0.35\textwidth
\centerline{\epsfbox{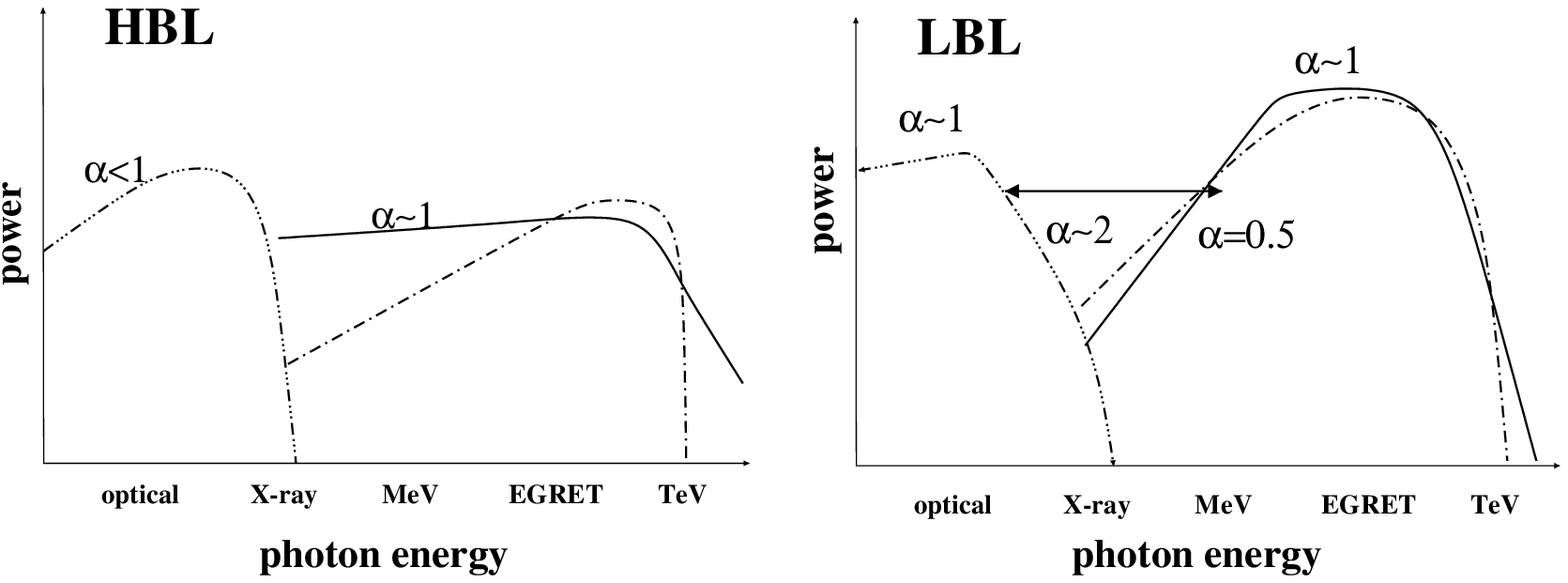}}
\figcap{Generic SS-PIC spectra for HBL (left) and LBL (right) type
blazars. Dashed high-energy lines show a typical spectral shape of an SSC
model for comparison.}
\text

\subsubsection*{Narrow cascades: proton and muon synchrotron radiation} 
\vspace{-5pt}

Apart from $\pi^0$ decay, energetic photons can also be produced by
synchrotron radiation of the protons. Moreover, it has been shown that also
the most energetic muons produced by the decay of photohadronically produced
charged mesons can lose a significant fraction of their energy in synchrotron
radiation in magnetic fields typical for SS-PIC blazar models, prior to their
decay \cite{RM98prd}. The total power ratio of synchrotron radiation
compared to the $\pi^0$ cascade injected by protons of the energy $E{\rm cr}$
in the observer's frame is 
\eqn{L:psyn:pi0}
\frac{L_{p, \rm syn}}{L_{\pi^0}} \sim 4 
	\left[\frac{E_{\rm cr}}{10^{20}\eV}\right]^{1-\alpha_t}
	\left[\frac{B}{10\G}\right]^2 
	\left[\frac{R}{10^{16}\cm}\right]^2 
	\left[\frac{\Gamma_{\rm jet}}{10}\right]^3 
	\left[\frac{L_{\rm IR}}{10^{44}\erg/\scnd}\right]^{-1}, 
\text
where $L_{\rm IR}$ is the observed infrared luminosity of the blazar. For
muon synchrotron radiation the ratio is $L_{\mu,\rm syn}/L_{\pi^0} \approx 2$
because of the photohadronic branching ratios \cite{sophia-texas}.

If the jet contains cosmic ray protons up to an energy $\hat E_{\rm cr}\sim
3\mal 10^{20}\eV$ in the observer's frame, $B\gsim 10\G$ and the typical
parameters assumed in \eq{L:psyn:pi0} otherwise, proton synchrotron radiation
extends up to observable energies of $B \hat x_p^2 m_e^3 \Gamma_{\rm jet}/(4
B_c m_p^3) \gsim 300\GeV$. Muon synchrotron radiation requires a minimal
dimensionless energy $x^* = E_\mu/m_e c^2 \Gamma_{\rm jet} \sim 5\mal 10^{11}$
for $B\sim 10\G$ \cite{RM98prd}, and therefore extends from $30\GeV$ to
$3\TeV$ using the same parameters. Since we have argued above that the
opacity break is in the same regime, a significant fraction of this radiation
can be reprocessed in a further cascade generation and would appear in the
hard X-ray regime up to energies of ${\sim}10\keV$ and ${\sim}100\keV$ for
reprocessed proton and muon synchrotron radiation, respectively.

\setfig{t}{narrow}
\epsfysize0.55\textwidth
\centerline{\epsfbox{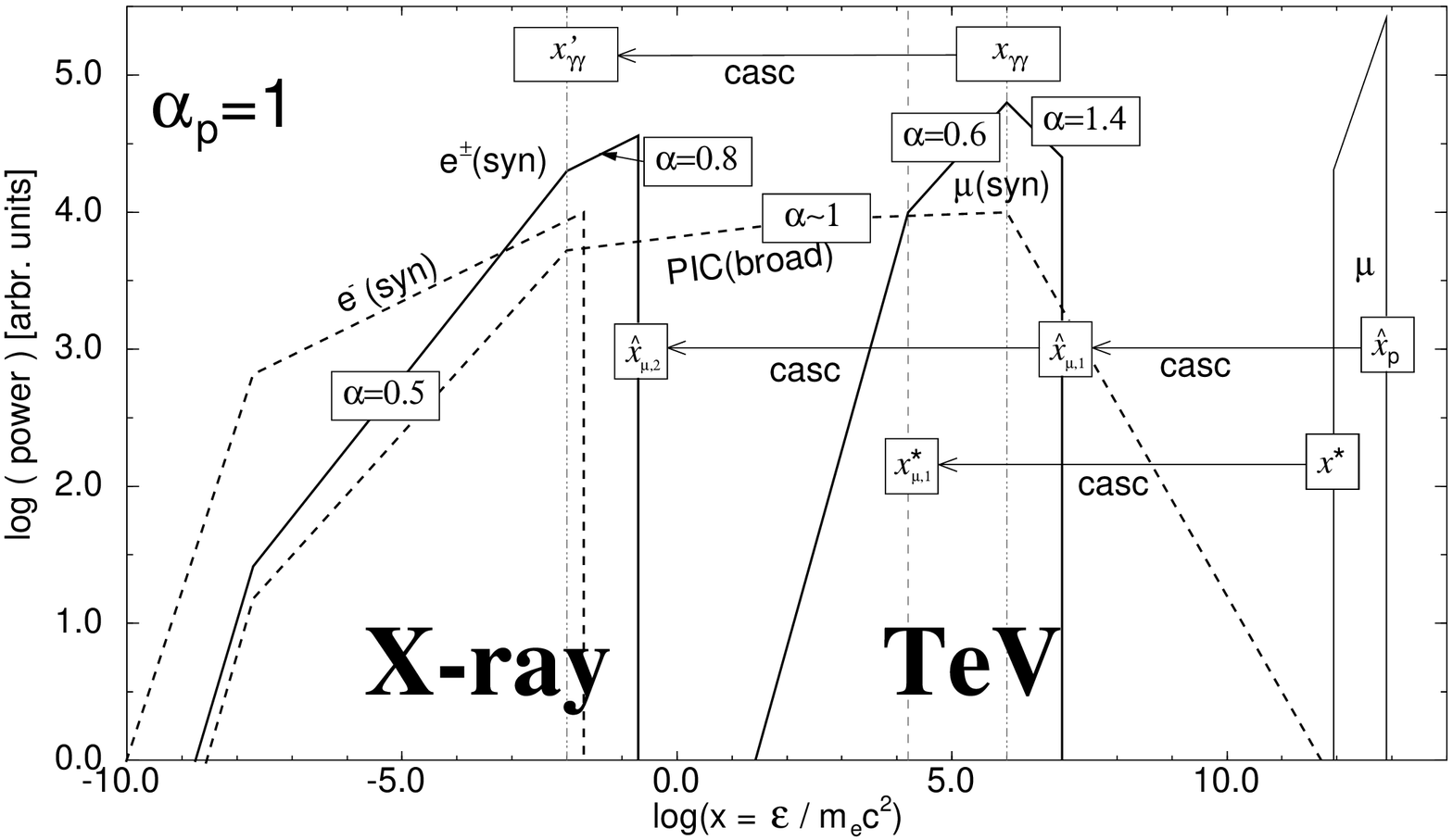}}
\figcap{Schematic spectra of narrow cascades superposing the broad band
emission of ordinary $\pi^0$ cascades and primary electron synchrotron
radiation --- Spectral indices correspond to $\mu$-induced cascades for
$\alpha_p = 1$, arrows indicate the cascading of spectral features. The
proton-synchrotron cascade is omitted for clarity, but has a similar
structure at photon energies about a factor of $10$ lower (see also
\cite{MP99Utah}).}
\text
The interesting aspect about these cascades becomes obvious when we consider
their spectrum below the opacity break. Here, we assume for simplicity that
proton synchrotron radiation is significant, but not dominant over adiabatic
or advection losses. In this case, the proton synchrotron spectrum has an
energy index $\alpha_{p,1} = \frac12 \alpha_p \sim 0.5$.  For muons
synchrotron cooling is always dominant (otherwise it is suppressed due to muon
decay), and we obtain $\alpha_{\mu,1} = f^-[\alpha_p] \sim 0.5$ for
$x_{\gamma\gamma} > x > x^*_1 = B {x^*}^2 m_e^3/(4 B_c m_\mu^3)$, and
$\alpha_{\mu,1} = -\frac13$ below from synchrotron emission below the
characteristic frequency.  Hence, both processes produce spectra very much
flatter than those arising from ordinary, $\pi^0$ induced cascades. Above the
opacity break, the spectra steepen by $\alpha_t\sim 1$, which means that the
peak of the luminosity is reached at the opacity break. Here, the dominance
of this component over the broad $\pi^0$ cascades is even stronger than
expected from the power ratio factors derived above, since it is first
generation synchrotron emission, while the $\pi^0$ emission from the same
protons is broadened through the cascading process. The spectral indices
expected from the narrow cascades below and above the opacity break
(\fig{narrow}) are consistent with the typical EGRET and TeV indices,
respectively, observed in Mrk\,421 and Mrk\,501 \cite{Sambruna}. The spectral
indices of the second cascade are obtained from a second application of the
operator $f^{-}$, yielding again values around $0.5$ steepening by $\frac12
\alpha_t$ above an X-ray break observable at $m_e c^2 \Gamma_{\rm jet}
x'_{\gamma\gamma}$ with $x'_{\gamma\gamma} \approx B x_{\gamma\gamma}^2/4
B_c$ (see \fig{narrow}). These compare well to the indices observed by
Beppo-SAX in three prominent flares of Mrk\,501 in April 1997
\cite{PVT+98}. The luminosity in the X-ray peak depends sensitively on the
energy of the opacity break, which determines how much energy of the TeV peak
is reprocessed. For example, in the SS-PIC model by M\"ucke and Protheroe
\cite{MP99Utah} the opacity break is at $m_e c^2 x_{\gamma\gamma} \Gamma_{\rm
jet} \sim 25\TeV$, which explains why their X-ray peak from reprocessed
proton synchrotron radiation is strongly suppressed.  For suitable parameters,
however, {\em narrow PIC emission can produce a two-bump spectrum with
comparable peaks the TeV and the X-ray regime}, as illustrated in Figure
\reffig{narrow}.

\section*{Variability and correlated flares}
\vspace{-3pt}
\subsubsection*{Flares and the quiescent background}
\vspace{-5pt}

A major result of the simultaneous multiwaveband observations of Mrk\,421 and
Mrk\,501 was that their variability in the X-ray and TeV-regime is largely
correlated, and stronger than in other wavebands. This has been used as an
argument against a hadronic interpretation of their gamma-ray emission, since
ordinary cascade models expect a quite model independent spectral index (as
explained above), which implies that the gamma-ray variability of such
sources should be comparable at all gamma-ray energies. 

Obviously, this picture changes if we consider narrow cascades. In order to
play a dominant role in the emission, these require conditions which allow to
produce cosmic rays of extremely high energies. Let us assume now that such
conditions are not always present in the jet, but only in some confined
regions for a limited time. Of course, we may still assume that also in other
regions the jet accelerates protons and electrons, albeit not to such high
energies. This would lead to a permanent ``glow'' of the jet, which is
dominated by the emission of primary electrons and $\pi^0$ induced cascades
with a spectrum comparable to the case illustrated in \fig{spectra} for HBLs,
which is consistent with observations of the quiescent emission of
Mrk\,421. A short flare which contains UHE protons able to produce narrow
cascades, and which is energetic enough to compete with the total emission of
the rest of the jet, would then cause the strongest variability in the energy
regimes where the flare spectrum peaks, and these are the X-ray and TeV
regimes. At other wavebands, the variability may be low, or not present at
all since the emission there may continue to be dominated by the quiescent
background (see \fig{flares}).
\setfig{t}{flares}
\epsfysize0.45\textwidth
\centerline{\epsfbox{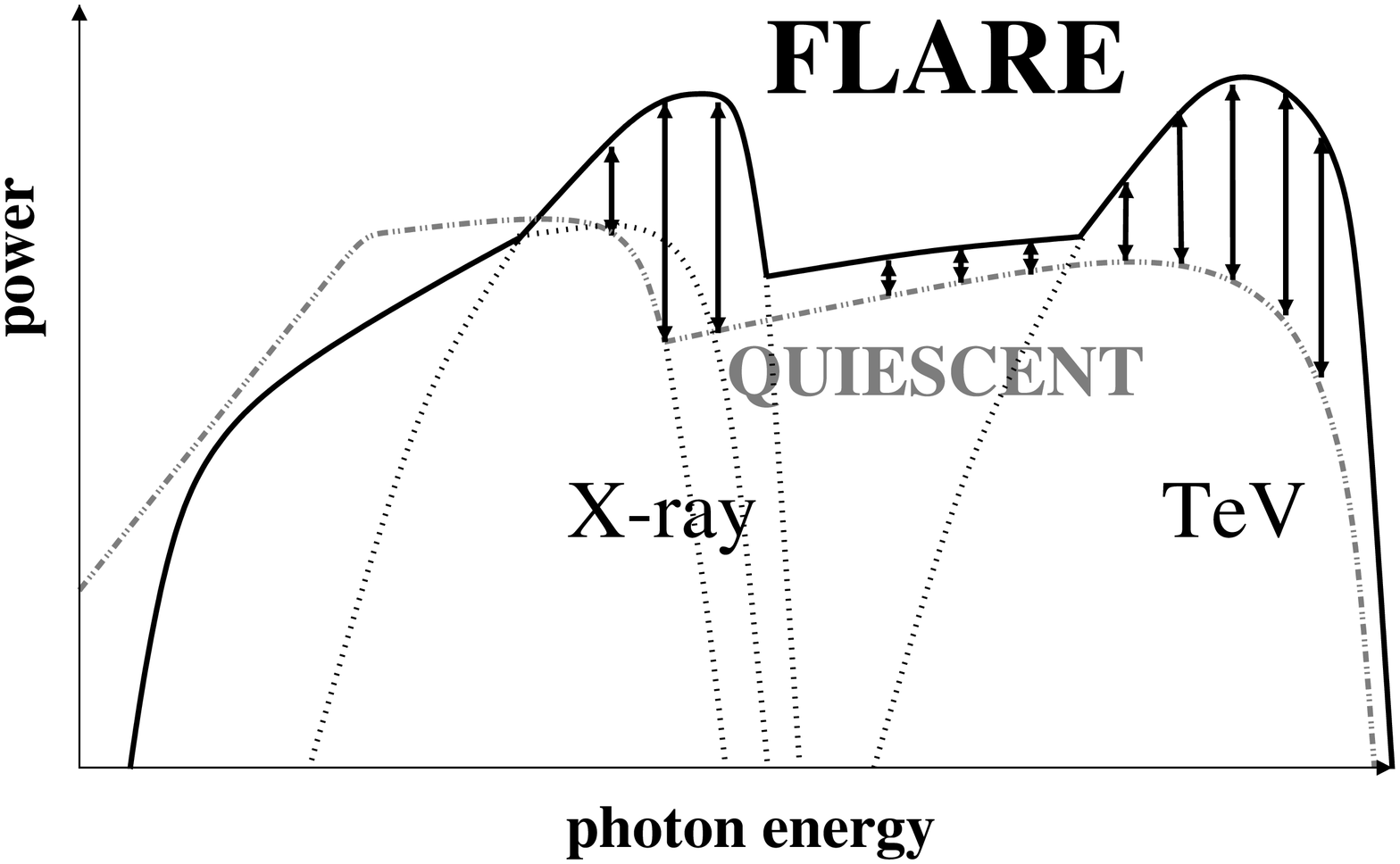}}
\figcap{Correlated variability on blazars as a result of flaring, 
narrow PIC emission superposing a non-peaked background from the surrounding
jet.} 
\text

\subsubsection*{Opacity and variability}
\vspace{-5pt}

The assumption of the existence of an opacity break in or below the TeV
regime is vital for a successful explanation of blazar spectra in an SS-PIC
model. This has some very interesting implications if the emission is
variable. Let us assume a nearly plane-parallel geometry of length $R$ for a
region emitting photons simultaneously over its entire volume, over a time
scale $t_{\rm rad} \ll R/c$. In the optically thin case, this would induce a
flare of duration $R/c$ owing to the run-time differences of photons. If the
emitter is opaque at some some energy $x$, i.e., $\tau_{\gamma\gamma}(x)\gg
1$, only photons emitted within a distance $R/\tau_{\gamma\gamma}$ can reach
the observer, causing a flare of duration $R/c\tau_{\gamma\gamma}$.

Applying this to SS-PIC models with an opacity break below TeV, we would
therefore expect that the observed multi-TeV variability becomes
systematically faster with increasing photon energy, while the variability
time scales are energy independent at X-ray or EGRET energies that are below
the opacity break. Of course, the simple relation $T_{\rm var} = T_{\rm
var,0}/\tau_{\gamma\gamma}$ for $\tau_{\gamma\gamma}>1$, and $T_{\rm var} =
T_{\rm var,0}$ for $\tau_{\gamma\gamma}<1$ is unlikely to apply to SS-PIC,
since both cascade injection and propagation involves time scales ${\sim}
R/c$. Nonetheless, the effect remains qualitatively in the sense that it
reduces the variability time from $\sim t_{\rm rad} + R/c$ to $\sim t_{\rm
rad}$ for $\tau_{\gamma\gamma}\gg 1$. This may explain some recent
simultaneous X-ray/TeV observations of Mrk\,501 which indicate a TeV
variability about a factor of $3$ shorter than at three different X-ray
energies, which all show the same time curve \cite{Sambruna}. In contrast,
optically thin SSC models would expect that there is some X-ray energy which
shows the same variability as observed in the TeV.

\section*{Supplementary aspects}
\vspace{-5pt}
\subsubsection*{Particle acceleration in the SS-PIC model}
\vspace{-5pt}

Some of the properties of the SS-PIC model discussed above on a purely
heuristical base (i.e., the assumption that UHECR are produced) can be put on
a more rigid foundation if we include the acceleration process. For jets the
most reasonable assumption is Fermi acceleration at shocks or plasma
turbulence \cite{Dru83rpp}.  If a particle with mass $m$ is accelerated on a
time scale $t_{\rm acc} = \eta^{-1} t_{\rm L}$, where $\eta$ has the same
meaning as in \eqs{Emax}, then the cutoff of the synchrotron radiation
emitted by the particle satisfies the condition
\eqn{Emaxsyn:acc}
\hat \eps \lsim \pi \eta\, m c^2/\alpha_{\rm f} \sim 400 \eta\, m c^2\;,
\text
which is reached if the particle energy limited by synchrotron cooling
($t_{\rm acc} = t_{\rm syn}$). Applying \eq{Emaxsyn:acc} to synchrotron
radiating protons we find a maximum observed photon energy of
${\approx}\,[4\TeV] \eta_p$ (assuming $\Gamma_{\rm jet}\sim 10$). For Fermi
acceleration we can write $\eta(E) \sim [\beta\,\delta B(r_{\rm L})/B]^2$,
where $\beta$ is the velocity of the shock or plasma wave, and $\delta
B(r_{\rm L})/B$ is the fractional magnetic field turbulence on the scale of
the particle gyro-radius, $r_{\rm L} = E/e B$ (see \cite[App.\,D]{RM98prd}, and
references therein). Obviously, in order to explain photons observed above
$300\GeV$ by proton synchrotron radiation we need $\eta(\hat E_{\rm cr}) >
0.1$, which requires relativistic shocks and strong turbulence on the largest
scales in the system. In this picture, correlated flares in blazars are due
to the appearance of transient relativistic shocks in a turbulent flow, while
the background emission is due to continuous acceleration at omnipresent weak
shocks or plasma turbulence.

Applying \eq{Emaxsyn:acc} to accelerated electrons, we find a maximum
synchrotron photon energy of ${\approx}\,[2\GeV] \eta_e$ for $\Gamma_{\rm
jet}\sim 10$, which implies $\eta_e \sim 10^{-9}{-}10^{-5}$ to explain the
observed IR-X cutoffs in blazars. The large difference between $\eta_p$ and
$\eta_e$ can be understood from the theory of plasma turbulence, noting that
the electrons probe very much smaller scales of the turbulence than the
protons. If the turbulence is described by $\eta(E) \propto [\delta B(r_{\rm
L})/B]^2 \propto E^y$, we find
\eqn{p:e:turb}
\hat\eps_{e,\rm syn} \sim  \hat\eps_{p,\rm syn} 
	\Big[m_e/m_p\Big]^{(3y+2)/(2-y)},
\text
which is obviously independent of Doppler boosting. Biermann and Strittmatter
\cite{BS87apj} pointed out that the near infrared cutoffs observed in many
quasars (e.g., LBLs) can be understood if they accelerate protons to the GZK
limit, and the plasma turbulence is described by a Kolmogorov spectrum
($y=\frac23$). In terms of \eq{p:e:turb}, we would obtain $\hat\eps_{e,\rm
syn}\sim 1\eV$ for $y=\frac23$ and $\hat\eps_{p,\rm syn} = 10\GeV$, which is
consistent with the observations of LBLs and would imply $\eta_p\sim
10^{-3}$. The Kolmogorov spectrum applies to fully developed hydrodynamical
turbulence if the magnetic field does not significantly contribute to the
total energy density of the fluid. In a magnetically dominated plasma a
Kraichnan turbulence spectrum would be expected \cite{BS87apj}, that is
$y=\frac12$. Combining this with $\hat\eps_{p,\rm syn} \sim 300\GeV$ which we
required for blazars with correlated X-ray/TeV variability, we find
$\hat\eps_{e,\rm syn} \sim 10\keV$, which corresponds to X-ray cutoffs
typically observed in HBLs (note that the flare emission up to ${>}100\keV$
in Mrk\,501 is explained by hadronic cascades in this model). This allows an
interesting explanation of the physical difference between these blazar
classes: while LBL jets are energetically dominated by the hydrodynamic
motion of the plasma, and involve only non-relativistic shocks and/or weak to
moderate turbulence, HBL jets are strongly turbulent, magnetically dominated
flows, in which also relativistic shocks occur. More aspects of particle
acceleration in the SS-PIC model are discussed in \cite{MP99Utah}.

\subsubsection*{Gamma rays, cosmic rays and neutrinos}
\vspace{-5pt}

One prediction which is unique to hadronic blazar models is the production of
energetic neutrinos with a luminosity comparable to gamma-rays. The detection
of such neutrinos, in particular if correlated with blazar flares, would thus
be a ``smoking gun'' for the hadronic scenario; unfortunately, the neutrino
fluxes from single blazer flares are that low that this can hardly be
expected within the next decades \cite{RM98prd}.  However, we would expect to
find {\em diffuse} VHE neutrino fluxes comparable to the the diffuse
extragalactic gamma ray background (DEGRB), if a considerable fraction of the
extragalactic gamma rays are of hadronic origin. This would also imply that a
significant fraction of the UHECR flux is produced by the same process
\cite{MPR99prd}. To utilize these relations to decide the nature of gamma-ray
emission in blazars, we need to determine which fraction to the DEGRB they
contribute --- which is again a task for gamma-ray astronomy.

In conclusion, although VHE neutrino observations will be very important to
clarify the origin of cosmic rays, I still see the currently better technical
possibilities to decide this question on the side of gamma-ray astronomy. To
do this by revealing the nature of gamma-ray emission from blazars will
require (i) a complete coverage of the gamma-ray wavebands from MeV to
multi-TeV energies with sufficient spectral resolution, together with more
campaigns allowing truly simultaneous multifrequency monitoring, and (ii) a
comprehensive discussion of the data in the view of possible leptonic {\em
and} hadronic explanations.

\vspace{1pc}

\begin{small}

{\bf Acknowledgments.} I wish to thank A.~Achterberg, A.~Atoyan, R.~Bingham,
J.~Kirk, K.~Mannheim, A.~Mastichiadis, A.~M\"ucke, and R.~Sambruna for
interesting and helpful discussions, and the LOC for support allowing me to
visit the meeting. This work was supported in part by the EU-TMR network
Astro-Plasma\,Physics, under contract number ERBFMRX-CT98-0168.

\end{small}
 

\begin{references}

\bibitem{Bie97jphG}
P.~L. Biermann, {\jphysg} {\bf 23},  1  (1997), and references therein.

\bibitem{Dru83rpp}
L.~{O'C}. {Drury}, {\reprophys} {\bf 46},  973  (1983), and references therein.

\bibitem{RM98prd}
J.~P. {Rachen} and P. {M{\'e}sz{\'a}ros}, {\physrevd} {\bf 58},  123005  (1998).

\bibitem{GZK66}
K. Greisen, {\prlett} {\bf 16},  748; G. Zatsepin and V. Kuzmin, {\JETPlett}
{\bf 4},  78  (1966). 

\bibitem{Ste68prl}
F.~W. {Stecker}, {\prlett} {\bf 21}, 1016 (1968).

\bibitem{LSSacc}
C.~A. {Norman} {\it et al.}, {\apj} {\bf 454},  60 (1995); H. {Kang} {\it et
al.}, {\mnras} {\bf 286},  257  (1997).

\bibitem{RB93aa}
J.~P. {Rachen} and P.~L. {Biermann}, {\aap} {\bf 272},  161  (1993).

\bibitem{Man93aa} K. {Mannheim}, {\aap} {\bf 269},  67  (1993); {\ssr} {\bf
75},  331  (1996). 

\bibitem{MPR99prd} K.~Mannheim {\it et al.}, {\physrevd} submitted,
astro-ph/9812398. 

\bibitem{GRBCR95} M. {Vietri}, {\apj} {\bf 453}, 883; E. Waxman, {\physrevl}
{\bf 75}, 386 (1995).

\bibitem{UPad95pasp}
C.~M. {Urry} and P. {Padovani}, {\pasp} {\bf 107},  803  (1995).

\bibitem{BBD+98aa}
N. {Bade} {\it et~al.}, {\aap} {\bf 334},  459  (1998).

\bibitem{MKB91aa} K. {Mannheim} {\it et al.}, {\aap} {\bf 251}, 723 (1991).

\bibitem{pp-junk} A. {Dar} and A. {Laor}, {\apjl} {\bf 478}, L5 (1997);
M. {Pohl}, these proceedings.

\bibitem{Apostolos} J.~G. {Kirk} and A. {Mastichiadis}, {\nat} {\bf 360},
135 (1992); D. {Kazanas} and A. {Mastichiadis}, {\apjl} {\bf 518}, L17 (1999).

\bibitem{Pro97iau} R. Protheroe, in {\em Accretion Phenomena and Related
Outflows}, ed. D. Wickramasinghe {\it et al.}, IAU Colloquium 163, p.~585
(1997).

\bibitem{PB97app} R. Protheroe and P.~L. Biermann, {\apph} {\bf 6}, 293
(1997). 

\bibitem{Sambruna} R. Sambruna, these proceedings, and references therein.

\bibitem{Sve87mnras} R.~Svensson, {\mnras} {\bf 227}, 403 (1987).

\bibitem{Man93prd} K.~Mannheim, {\physrevd} {\bf 48}, 2408 (1993).

\bibitem{sophia-texas} A. M{\"u}cke {\it et~al.}, Proc.~19th Texas
Symposium, Paris, 1998, astro-ph/9905153.

\bibitem{PVT+98}
E. Pian {\it et~al.}, {\apjl} {\bf 497},  L17  (1998).

\bibitem{MP99Utah} A. M\"ucke and R. Protheroe, these proceedings, and
private communication.

\bibitem{BS87apj} P.~L. {Biermann} and P.~A. {Strittmatter}, {\apj} {\bf
322}, 643 (1987), and references therein.


\end{references}
\end{document}